\documentclass[twocolumn,prd,superscriptaddress,showpacs,nofootinbib]{revtex4}

\usepackage{graphicx}
\usepackage{dcolumn}
\usepackage{bm}
\usepackage{hyperref}

\begin{document}


\title{Detection potential for the diffuse supernova neutrino background in the large liquid-scintillator detector LENA}

\author{M. Wurm}\email[Corresponding author, e-mail:~]{mwurm@ph.tum.de}
\author{F. von Feilitzsch}
\author{M. G\"oger-Neff}
\affiliation{Physik-Department E15, Technische Universit\"at M\"unchen, James-Franck-Str., D-85748 Garching, Germany}
\author{K. A. Hochmuth}
\affiliation{Max-Planck-Institut f\"ur Physik, F\"ohringer Ring 6, D-80805 M\"unchen, Germany}
\author{T. \surname{Marrod\'an Undagoitia}}
\author{L. Oberauer}
\author{W. Potzel}
\affiliation{Physik-Department E15, Technische Universit\"at M\"unchen, James-Franck-Str., D-85748 Garching, Germany}

\date{\today}

\begin{abstract}

The large-volume liquid-scintillator detector LENA (Low Energy Neutrino Astronomy) will provide high-grade background discrimination and enable the detection of diffuse supernova neutrinos (DSN) in an almost background-free energy window from $\sim10$ to 25\,MeV. Within ten years of exposure, it will be possible to derive significant constraints on both core-collapse supernova models and the supernova rate in the near universe up to redshifts $z<2$.

\end{abstract}



\maketitle

\section{Introduction}

The cosmic background of neutrinos generated by core-collapse supernova explosions throughout the universe is known as supernova relic neutrinos (SRN) \cite{ando0401} or more precisely as diffuse supernova neutrinos (DSN) \cite{sk-dsn-06}. These neutrinos are generally believed to provide a new source of information on the core-collapse supernova explosion mechanism and both on the supernova rate (SNR) and on the star formation rate (SFR) up to high redshifts of $z\simeq5$ \cite{ando0401,sfr-05,lunardini0610}.\\
However, the detection of DSN is demanding, as both their integral flux of $\sim 10^2\,\nu\,\text{cm}^{-2}\text{s}^{-1}$ and their energy of $E_\nu<50\,$MeV are low. In general, all $\nu$ and $\bar\nu$ flavours are contained in the DSN. However, $\bar{\nu}_{e}$ are the most likely to be detected as the inverse beta decay reaction, $\bar\nu_{e}+p\to e^{+} + n$, has the largest cross section in the DSN's energy region \cite{ando0401}. Liquid scintillator detectors (LSD) provide a large number density of free protons, a clear delayed coincidence signal for this detection channel and high energy resolution at low energies. A large-volume LSD is therefore an ideal candidate for DSN detection.\\
The best experimental limit on the DSN flux $\Phi_{\bar\nu_e} < 1.2\,\text{cm}^{-2}\text{s}^{-1}$ for E$_{\bar{\nu}_e} > 19.3$\,MeV (90\,$\%$ C.L.) has been achieved by the Super-Kamiokande experiment \cite{lunardini0610,SK-2003-SRNlimit}. However, this limit is strongly determined by the background events present in a pure water Cherenkov detector (WCD). As a liquid-scintillator detector (LSD) allows far better background discrimination, it opens a nearly background-free energy window from $\sim10$ to 25\,MeV for DSN detection. The target masses of KamLAND \cite{kamland-2004} and of near-future experiments like BOREXINO \cite{borex-sn} and SNO+ \cite{sno+dsnb} are not sufficient to reach significant statistics \cite{ando-2003-18}. A large-scale detector like LENA (Low Energy Neutrino Astronomy) \cite{LENA-05,LENA-06} with about 50\,kt of liquid scintillator is required.\\
We show in the present paper that LENA allows the detection of $\sim10$ DSN events per year in the almost background-free energy range from about 10 to 25\,MeV. After an observation time of $\sim10$ years a spectral analysis of the DSN is possible, having implications on both core-collapse supernova (SN) models and the SNR up to redshifts $z\simeq2$. If no signal was detected, the new limits are significantly lower than all model predictions and improve the limit given by Super-Kamiokande by a factor of $\sim9$.\\
This paper mainly focusses on the observational window that is given by the remaining background of atmospheric and reactor $\bar{\nu}_{e}$ that are indistinguishable from the actual DSN signal. In Sections~\ref{LENA} and \ref{theory}, a description of the planned detector and an overview of the predictions on the DSN flux and spectrum are given. In Sect.\,\ref{detection} the $\bar\nu_e$ detection channel is discussed. The results of the calculations on the site-dependent reactor neutrino background rate and its spectral shape, considering especially the highly energetic part, are shown in Sect.\,\ref{background}. The cosmogenic background, mainly $^9$Li and fast neutrons produced by muons, is shortly discussed in Sect.\,\ref{muons}.  Using this information, the expected event rates for LENA inside the energy window from $\sim10$ to 25\,MeV are given for different DSN models in Sect.\,\ref{discpot}. Furthermore, the spectral discrimination of different DSN models has been investigated by Monte Carlo simulations. Their results as well as a method to find a lower boundary for the SNR at $\text{z}<2$ are also described in this section. Conclusions are given in Sect.\,\ref{conclusions}.

\section{Detector Concept}
\label{LENA}

\begin{figure}[h]
  \begin{center}
    \includegraphics[width=0.4\textwidth]{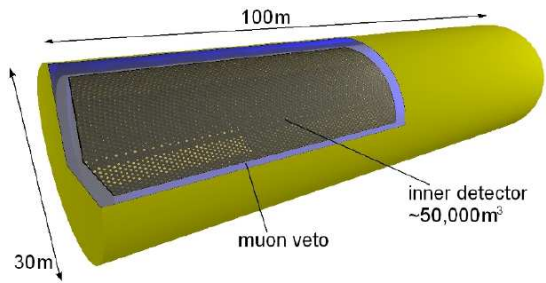}
    \caption{Sketch of the LENA detector\label{LENApic}} 
  \end{center}	
\end{figure}

The LENA concept foresees a cylindrically shaped detector of about 100\,m in length and 30\,m in diameter (see Fig.\,\ref{LENApic}). An inner part of 26\,m diameter contains about $5\times10^{4}\,\text{m}^3$ of liquid scintillator, whereas an outer mantle with a thickness of 2\,m filled with water is intended as a Cherenkov muon veto. The fiducial volume is defined within a radius of 12\,m of the inner cylinder and encloses a volume of $4.4\times10^{4}\,\text{m}^3$. About 12\,000 photomultipliers (PMs, 50\,cm diameter) installed on the inner surface provide a surface coverage of about 30\,$\%$. The effective coverage could be further increased by mounting light-collecting concentrators to the PMs.\\
At present, a preference exists for a liquid scintillator based on the organic solvent PXE (C$_{16}$H$_{18}$, $\rho = 985\,\text{g}/\ell$) that was already tested in the BOREXINO prototype, the Counting Test Facility (CTF) at the Gran Sasso National Laboratory (LNGS) \cite{borexcoll-2004-1}. Due to the large detector radius, an attenuation length of $\sim10$\,m at 430\,nm has to be achieved in order to obtain an adequate photoelectron (pe) yield in the PMs. It has been shown that this aim can be reached by purification of the PXE in an Aluminum-column \cite{borexcoll-2004-1,mydipl}. Furthermore, adding 80 weight-percent of Dodecane (C$_{12}$H$_{26}$, $\rho = 749\,\text{g}/\ell$) to the solvent has a positive impact on the transparency of the liquid and increases the number of free protons in the scintillator - and therefore the $\bar\nu_e$ event rate - by almost $25\,\%$ \cite{mydipl}. As Dodecane slightly lowers the light yield of the scintillator, both pure PXE and the described mixture provide similar photoelectron yields of at least 100\,pe/MeV for an event in the center of the detector. As primary and secondary wavelength shifters (fluors) 6\,g/$\ell$ PPO and 20\,mg/$\ell$ bisMSB will be used \cite{mydipl}.\\
Besides detection of the DSN $\bar\nu_e$, LENA will be an observatory for solar neutrinos \cite{LENA-05}, geoneutrinos \cite{hochmuth05}, and the neutrinos emitted by a galactic core-collapse SN \cite{LENA-05,LENA-06,Oberauer0402,Skadhauge0611}. Moreover, the detector will allow to investigate the properties of neutrinos via beam experiments, and to search for proton decay \cite{teresa-2005-1, Marrodan06TAUP}.\\
At the moment, the preferred detector sites are a mine, the Center of Underground Physics in Pyh\"asalmi (CUPP, Finland) \cite{peltoniemi0506}, or the underwater plateau in the Mediterranean Sea used by the NESTOR Collaboration next to Pylos (Gree\-ce). Both sites provide an effective shielding of $\sim4000$\,m.w.e. against cosmic radiation. We have chosen LENA at CUPP as our default scenario, as Pyh\"asalmi is far away from the middle-European reactors and is able to provide the required infrastructure for a large-scale experiment. In addition, local representatives have signaled big interest in the project. However, as the background due to power reactor $\bar{\nu}_{e}$-flux is most important, locations in France, the US, next to the islands of Hawaii, and New Zealand have also been investigated for comparison (see Sect.\,\ref{background}).

\section{Theoretical Predictions}
\label{theory}

The energy-dependent DSN flux discussed in this paper was taken from the publications by S. Ando \textit{et al.} \cite{ando0401,ando0402}. In general, theoretical models for the expected DSN spectrum depend on two sources of input:\\

\textbf{Supernova neutrino spectrum.} The spectral form of the DSN flux is strongly dependent on the supernova (SN) core-collapse model applied. Three different predictions have been made in \cite{ando0401}\ with reference to the SN simulations performed by the Lawrence Livermore Group (LL) \cite{LL-1998}, by Thompson, Burrows and Pinto (TBP) \cite{TBP-2003} and by Keil, Raffelt and Janka (KRJ) \cite{KRJ-2003}.\\
The observed $\nu$ spectra on earth will differ from those emitted by the proto-neutron star as the $\nu$ and $\bar\nu$ are affected by the matter potential of the dying star \cite{dighe9907,Kotake0509}. As the neutrinos pass from the large matter potential of the central region to the surrounding potential-free vacuum, the change in the mixing parameters leads to a partial conversion of $^(\bar\nu^)_e$ into $^(\bar\nu^)_{\mu,\tau}$ and vice versa. In most SN models, the larger mean energy of $^(\bar\nu^)_{\mu,\tau}$ at production hardens the resulting $^(\bar\nu^)_e$ spectrum. In addition, all models predict at least one resonant flavour conversion inside the SN envelope due to the mixing angle $\theta_{12}$ that applies to the neutrino sector only. A second resonance might occur if the value of $\theta_{13}$ is larger than $\sim1^\circ$ and therefore enables an adiabatic conversion. This will affect either the $\nu$ sector in case of a normal neutrino mass hierarchy, or the $\bar\nu$ for an inverted hierarchy. In the case of full adiabaticity the resonance will lead to an almost complete conversion of $\bar\nu_e \leftrightarrow \bar\nu_{\tau}$ as $\theta_{13}$ is limited to values $\theta_{13}\leq12.5^{\circ}$ \cite{dighe9907,Kotake0509,Chooz9711}. In the following, we will refer to this resonant conversion of $\bar\nu_e$ and $\bar\nu_{\tau}$ as \textit{resonance} case. All other combinations of the value of $\theta_{13}$ and the mass hierarchy result in identical $\bar\nu_e$ spectra. This situation we call \textit{no resonance} case.\\
If the mean energy of the $\bar\nu_{\tau}$ is much larger than that of the $\bar\nu_e$ as predicted by the LL model, the \textit{resonance} case noticably shifts the spectrum of the $\bar\nu_e$ to higher energies. However, if one assumes only a small gap between those energies as in the KRJ model, the effect becomes negligible. For further explanation see \cite{ando0401,dighe9907,Kotake0509}.\\

\textbf{Supernova rate.} Additional information is required on the redshift-dependent supernova rate $\dot\rho_{SN}(z)$ (SNR)\footnote{In the present paper, SNR is always referring to the rate of the core-collapse SN of type Ib,c and II.}. The SNR can either directly be measured \cite{cappellaro9904,dahlen0406} or can be derived by combining the observed star formation rate $\dot\rho_*(z)$ (SFR) with the initial mass function (IMF) of the forming stars \cite{lunardini0509}. As only massive stars with more than 8 solar masses (M$_{\odot}$) will end in a core-collapse SN \cite{ando0401}, their rather short lifecycles constitute a negligible aberration in comparison to cosmic time scales. Nevertheless, there is an uncertainty in the proportion of heavy SN-progenitor stars due to the uncertainty in the mass dependence of the IMF \cite{lunardini0509,hopkins0601} at low masses. The standard Salpeter IMF ($\phi(m)\propto m^{-2.35}$) returns the relation $\dot\rho_{SN}(z)=0.0122\,M_\odot^{-1}\,\dot\rho_*(z)$ \cite{ando0401}, whereas for modified IMF assumptions the conversion factor varies from $-25\,\%$ to $+8\,\%$ relative to the Salpeter result \cite{hopkins0601}.\\

\begin{figure}
  \begin{center}
    \includegraphics[width=0.4\textwidth]{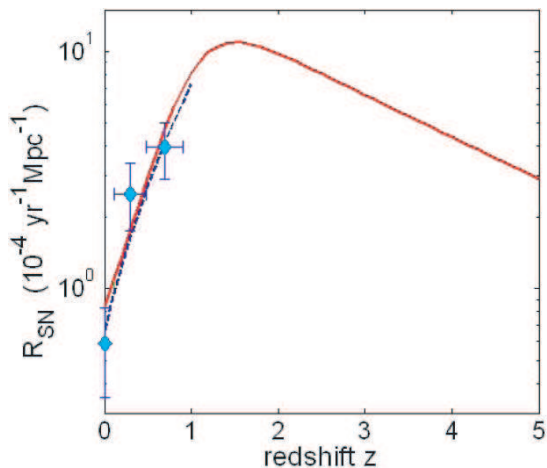}
    \caption{The z-dependent supernova rate calculated according to Ando \textit{et al.} \cite{ando0401} (\textit{solid line}) in comparison to the best fit ($dashed~line$) \cite{lunardini0509} of direct SNR observations (\textit{diamonds}) for $z<1$ \cite{cappellaro9904,dahlen0406}.\label{SNR}} 
  \end{center}	
\end{figure}

\textbf{Resulting DSN spectra.} In the present paper, we follow the model calculations by Ando \cite{ando0401} in combining an observationally obtained SFR with the Salpeter IMF. We obtain a redshift-dependent SNR

\begin{eqnarray}\label{SFRf}
\dot\rho_{SN}(z)= 3.9\times10^{-4}\,f_{SN}\,h_{70} \frac{e^{3.4z}}{e^{3.8z}+45}\nonumber\\ \times\frac{\sqrt{\Omega_m(1+z)^3+\Omega_\Lambda}}{(1+z)^{3/2}}\,\text{yr}^{-1}\,\text{Mpc}^{-3},
\end{eqnarray}

where $h_{70}=1$ for a Hubble constant of $H=70\,\frac{km}{s\,Mpc}$. $\Omega_m=0.3$ and $\Omega_\Lambda=0.7$ are the cosmological parameters of the  matter and dark energy density, respectively. The overall normalization uncertainty in the SNR is parameterised by $f_{SN}$. The resulting values of the SNR for $f_{SN}=1$ are in reasonable agreement with direct SN observations for $z<0.9$ \cite{lunardini0509,cappellaro9904,dahlen0406}, as shown in Fig.\,\ref{SNR}.

\begin{figure}
  \begin{center}
    \includegraphics[width=0.4\textwidth]{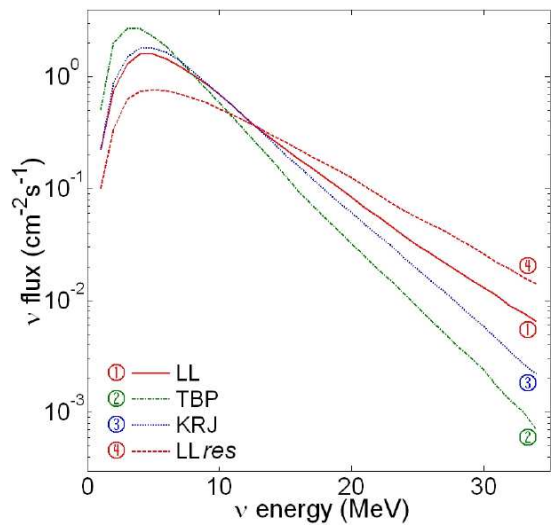}
    \caption{DSN model spectra for $\bar\nu_e$ calculated by Ando $et~al.$ \cite{ando0402} using core-collapse SN simulations performed by the Lawrence Livermore group (LL) \cite{LL-1998}, Thompson, Burrows and Pinto (TBP) \cite{TBP-2003} and Keil, Raffelt and Janka (KRJ) \cite{KRJ-2003}. In addition, an LL\textit{res} spectrum is shown in case of a resonant flavour conversion in the SN envelope. \label{SRNspec}} 
  \end{center}	
\end{figure}

In combination with the different SN models mentioned above one obtains three models for the DSN spectrum as depicted in Fig.\,\ref{SRNspec}, assuming $f_{SN}=1$ and the absence of a matter resonance in the $\bar\nu$ sector. For the LL model, an additional DSN spectrum LL\textit{res} is shown for the \textit{resonance} case.

\begin{figure}
  \begin{center}
    \includegraphics[width=0.4\textwidth]{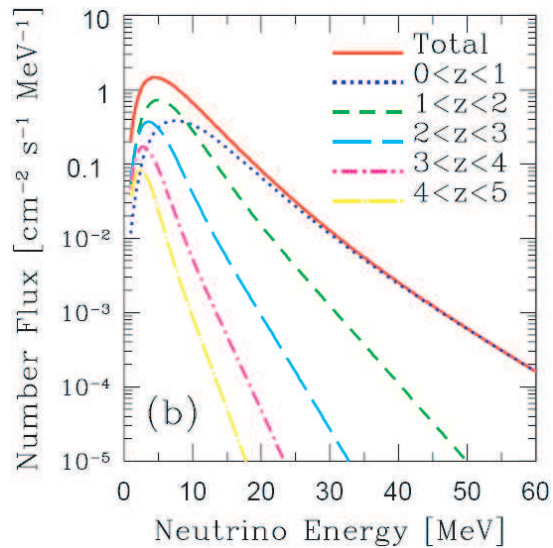}
    \caption{Contributions of different redshift regions to the DSN energy spectrum \cite{ando0402}. Neutrinos emitted by far away supernovae are redshifted by cosmic expansion.  \label{SRNz}} 
  \end{center}	
\end{figure}

Unlike the neutrinos that are produced by nearby SN the spectral portions corresponding to SN explosions further away will be increasingly redshifted by cosmic expansion \cite{ando0402}. For this reason, DSN from high-z regions will dominate the low-energetic part of the observed spectrum. As shown in Fig.\,\ref{SRNz}, the flux of neutrinos from sources at $z>1$ is substantially limited to energies E$_{\bar{\nu}_{e}} < 10\,$MeV.

As discussed above, the actual flux of DSN could vary widely from the predictions, as apart from the dependence on the used SN model there are  uncertainties due to the assumed value of $\dot\rho_{SN}(z)$. Direct measurements reach no further than redshifts $z=0.9$ and suffer from light extinction due to interstellar dust \cite{lunardini0509}. The same is true for observations of star formation regions in the ultraviolet (UV) \cite{UV-1,UV-2,UV-3} and far infrared (FIR) \cite{FIR-1,FIR-2} band. According to \cite{ando0401}, these uncertainties in the SFR translate to a factor $f_{SN}$ between 0.7 and 4.1 if one assumes a Salpeter IMF. However, combining the data of these observations with the cosmic gamma-ray background and the limits on the DSN flux by Super-Kamiokande, Strigari \textit{et al.} have derived a concordance model of the star formation rate (CMSFR) with a favoured value of $f_{SN}\simeq2.5$ \cite{strigari05}. New measurements of the SFR further reduce the parameter space, as discussed by Hopkins and Beacom \cite{hopkins0601}. While these new predictions develop differently with redshift especially for $z>1$, they are well contained inside the limits of $0.7<f_{SN}<4.1$ given by Ando \cite{ando0401}.\\
Current UV and FIR astronomy and their future projects might be able to determine both the SNR and the SFR with high accuracy \cite{lunardini0610}. The present uncertainty of the SFR is only $\approx30-50\,\%$ in the redshift region $z<1$ \cite{hopkins0601}. However, as the DSN detection is not suffering from dust extinction, neutrinos could test the validity of the assumed models for light extinction. Especially measurements in the energy region below 10\,MeV would provide valuable information on high redshift regions. However, in this energy regime reactor $\bar{\nu}_{e}$ prove to be an undistinguishable background and partially hide the DSN due to the high $\bar{\nu}_{e}$ rates, as will be discussed in Sect.\,\ref{background}.

\section{\texorpdfstring{$\bar\nu_e$ Detection Channel}{barnue detection channel}}
\label{detection}

As mentioned before, $\bar\nu_e$ are the best choice for the detection of the DSN background.
The cross section of the charged current interaction of a $\bar\nu_e$ with a Hydrogen nucleus in the target, the inverse beta decay $\bar\nu_e + p \to n + e^{+}$, is substantially larger than that of all other detection channels, $\sigma=6.8\times10^{-42}\,\textnormal{cm}^2$ at 10\,MeV \cite{Strumia0302}. Moreover, the reaction provides a low energy threshold of 1.8\,MeV (corresponding to the mass difference between proton and neutron plus positron).\\
Both decay particles can be detected in a liquid-scintillator detector, providing a coincidence that can be used for a very effective background reduction (see Sect.\,\ref{background}): The prompt signal due to the ionising processes and annihilation of the positron is followed by the signal of a 2.2\,MeV gamma quantum. This is released after a delay of $\sim 180\,\mu$s when the neutron is captured by a free proton in the scintillator, $n+p\to d+\gamma$.\\
Due to the large mass of the neutron the positron will carry most of the energy that was deposited by the $\bar\nu_e$. In good approximation, the kinetic energy of the positron will be reduced by $\sim1.8$\,MeV compared to the incoming neutrino due to the Q-value of the reaction. However, the annihilation of the positron in the scintillator adds another $2m_ec^2$ of energy to the detected signal, leading to a total reduction of $\sim0.8$\,MeV.\\
For the calculation of the expected event rate in LENA, the DSN flux must be convoluted with the energy-dependent cross section. The rate is then multiplied with the number of free protons in the fiducial volume. This number is $2.9\times 10^{33}$ for the case of a scintillator mixture of 20\,$\%$ PXE and 80\,$\%$ Dodecane and a fiducial volume of $44\times10^3\,\textnormal{m}^{3}$.\\
We calculated the expected energy resolution in LENA by using the results of both laboratory measurements of the scintillator properties and Monte Carlo (MC) simulations: The proposed scintillator mixture provides a light yield of $\sim8\times10^3$ photons per MeV, an attenuation length of $\sim10$\,m and a scattering length of $\sim30$\,m for the scintillation light (at 430\,nm) \cite{mydipl}. Assuming a wall coverage of $30\,\%$ and a quantum efficiency of $20\,\%$ for the PMTs, a photoelectron (pe) yield of at least 110\,pe/MeV can be achieved for an event in the center of the detector \cite{teresa-2005-1}. The resulting $1\sigma$ energy resolution is 0.10/$\sqrt{E}$ (in MeV).\\
For the three DSN models, we find in the \textit{no\,resonance} case that LL \cite{LL-1998} provides the largest event number: $\sim 6.8\,f_{SN} $ (see Eq.\,(\ref{SFRf})) detected $\bar\nu_e$ in 1 year of measurement time in LENA, followed by the KRJ ($\sim 6.1\,f_{SN}$) and by the TBP model ($\sim 4.7\,f_{SN}$). This variation of rates originates from the different spectral forms of the models: As the cross section increases with energy, the LL model which predicts the largest flux above 10\,MeV also provides the largest event rates. This is especially true for LL\textit{res} ($\sim 7.7\,f_{SN}$). For all models, most of the $\bar\nu_e$ events are expected to be in the energy region between 6 and 14\,MeV. However, for these estimates background events have been neglected.

\section{\texorpdfstring{$\bar\nu_e$ Background}{barnue background}}
\label{background}

A large fraction of the $\bar\nu_e$ events cannot be attributed to the DSN because of the background events due to $\bar{\nu}_{e}$ generated both by air showers and nuclear power plants. These are intrinsically indistinguishable from the $\bar{\nu}_{e}$ of the DSN and therefore independent of the type of detector used.\\

\subsection{\texorpdfstring{Reactor $\bar\nu_e$}{reactor barnue}}
\label{reactor}

For energies below $\sim10$\,MeV, the man-made background due to nuclear reactors sets a threshold for DSN detection. The $\bar{\nu}_{e}$ are generated by the $\beta^{-}$ decay of neutron-rich fission products of $^{235}$U, $^{238}$U, $^{239}$Pu and $^{241}$Pu inside the reactor. As the reactor $\bar{\nu}_{e}$ flux is quadratically declining with distance,
at least the closest reactors must be considered for a good estimate of the actual flux in the detector.\\
Above the Q-value of 1.8\,MeV, the spectral shape of the reactor neutrinos is best known up to energies of E$_{\bar\nu_e} = 8$\,MeV, both from experimental data and from theoretical calculations that have been using the fission yields and $\beta$ decay schemes of the isotopes in question. However, as the DSN flux is several orders of magnitude lower than that of the reactor neutrinos, it is necessary to take the high-energetic tail of their spectrum (up to E$_{\bar\nu_e}\approx 13$\,MeV) into account when determining the lower detection threshold for DSN observation.\\ 

\textbf{Reactor Neutrino Spectrum.} Especially for energies above 8\,MeV, the exact spectrum of the reactor $\bar{\nu}_{e}$ cannot be measured directly at a reactor because of the poor statistics. Instead, the spectrum is deduced from fission yields, $\beta$ endpoint energies and decay schemes of the neutron-rich isotopes produced in a reactor. The experimental challenge is set by the extremely neutron-rich isotopes with high Q-values and lifetimes in the range of 10$^{-2}$ seconds.\\
Tengblad \textit{et al.} have indirectly determined the reactor neutrino spectrum up to an energy of 12\,MeV by a measurement of the beta-decay spectra of the relevant fission products of $^{235}$U, $^{238}$U, $^{239}$Pu \cite{tengblad-1989}. These three elements are, on average, responsible for about 92\,$\%$  of the fission processes (see Table\,\ref{upu}) and therefore of the neutrino flux generated by the reactor \cite{fukugita}. In contrast to $^{238}$U, $^{241}$Pu contributes only a small portion of the fission products emitting high energetic neutrinos. For this reason, the spectral contribution of $^{241}$Pu can be neglected to good approximation.\\
However, there is at least one additional element known to be produced in fission processes which has an even higher $\beta$ endpoint energy: $^{94}$Br with a Q-value of 13.3\,MeV \cite{LBNLwww}. Due to its short lifetime of 70\,ms and its low fission yields (as shown in Table\,\ref{upu}) its exact $\beta$ decay scheme is not known \cite{letourneau-2005-talk}. Thus, in the present paper we have only been able to give an upper limit for its contribution to the spectrum, using its fission yield, Q-value, and the information that it decays in 70\,$\%$ of all cases without emitting an additional neutron \cite{letourneau-2005-talk}.\\

\begin{table}
	\caption{Contributions of the fission products of uranium and plutonium to the fission processes and therefore to 
	the $\bar{\nu}_{e}$ flux emitted by a reactor (averaged over time) \cite{fukugita}. The last column shows the fission yields of 	
	the high-endpoint $\beta$-emitter $^{94}$Br \cite{LBNLwww}.}\label{upu}
	\begin{ruledtabular}
		\begin{tabular}{lcc}
								&  Contribution to      &   $^{94}$Br       \\
		Isotope		 	&	 total fission rate  & 	fission yield					\\
		\hline
		$^{235}$U	 	&	 0.59  					& 	$1.66\times 10^{-6}$ 		\\
		$^{238}$U	 	&	 0.04 					& 	$7.90\times 10^{-5}$	 	\\
		$^{239}$Pu 	&	 0.285  				&		$2.71\times 10^{-5}$ 		\\
		$^{241}$Pu 	&	 0.075  				&		$1.05\times 10^{-6}$ 		\\
		\end{tabular}
	\end{ruledtabular}	
\end{table}

\begin{figure}
  \begin{center}
    \includegraphics[width=0.48\textwidth]{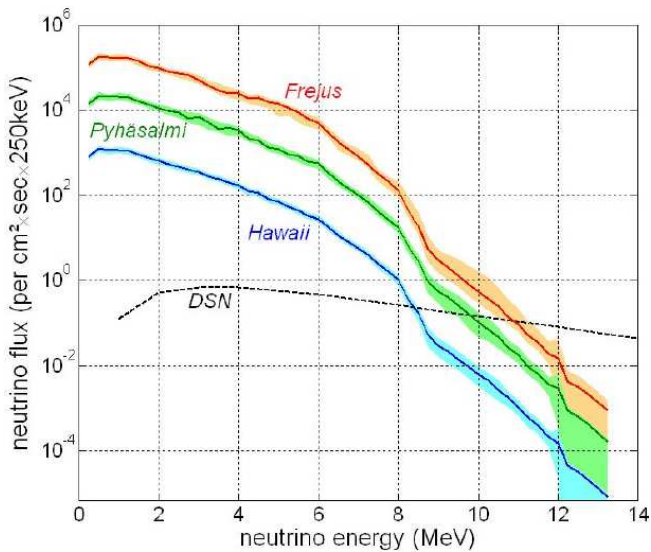}
    \caption{Spectra of reactor $\bar{\nu}_{e}$ at Frejus, Pyh\"asalmi and Hawaii. Shaded regions correspond to experimental and model uncertainties. Above 12 MeV, we added an upper limit for the spectral contribution of $^{94}$Br neutrinos. For comparison, the DSN spectrum according to the KRJ model \cite{KRJ-2003} with $f_{SN}=1$ (see Eq.\,(\ref{SFRf})) is shown. \label{3reactors}} 
  \end{center}	
\end{figure}

\textbf{Site-dependent Reactor Neutrino Flux.} Using the reactor $\bar\nu_e$ spectrum just described, the reactor background flux and the event rates in LENA were calculated at a number of different locations: At present Pyh\"asalmi (Finland) and Pylos (Greece) are the preferred detector sites. The Laboratoire Souterrain de Modane (LSM) at Frejus (France) was included as an example for a middle-European detector site. The US American sites Kimballton, Henderson, and Homestake Mine were used as they have uttered interest in a LENA-like liquid-scintillator detector and one of them will most likely be the home of the Deep Underground Science and Engineering Laboratory (DUSEL). Hawaii and New Zealand were chosen as they are far away from the nuclear power plants on the northern hemisphere and are from this point of view optimal detector sites for observing the DSN.\\
When calculating the $\bar\nu_e$ flux, in a first step the number of neutrinos emitted per second $R_{\bar\nu_e}$ by a nuclear reactor can be derived:
\begin{equation}\label{Nnu}
R_{\bar\nu_e} = N_{\nu,fiss} \frac{P_{th}}{E_{fiss}} e = (1.38\pm0.14)\,10^{20}  P_{th\,[GW]}\,\text{s}^{-1}, 
\end{equation}
where $N_{\nu,fiss} \simeq 6$ \cite{Zacek86} and $E_{fiss} = 205.3\pm0.6\,$MeV \cite{kopeikin-2003} are the average number of neutrinos and the mean energy produced per fission, respectively, P$_{th}$ is the thermal power of the reactor and $e = 0.75\pm0.06$ \cite{IAEAwww} is the average fraction of time that a reactor is running (energy availability factor $e$). The thermal power $P_{th}$ can be found in the online databases of the International Atomic Energy Agency (IAEA) \cite{IAEAwww}.\\
In a next step, using the coordinates of each reactor and the detector site their distance $d$ can be found. The flux without oscillations can be calculated dividing $R_{\bar\nu_e}$ by $4\pi d^{2}$.\\
Finally, for each distance to a particular power plant the effect of oscillations $\bar\nu_e \to \bar{\nu}_{\mu,\tau}$ has to be taken into account. This was achieved by convoluting the normalized spectrum $F(E)$ with the energy-dependent oscillation probability and weighing it with the expected integral flux,
\begin{equation}
\Phi(E,d)\textnormal{d}E =  \frac{ R_{\bar\nu_e}}{ 4\pi d^{2} }  F(E) (1-\sin^{2}(2\theta_{12})\sin^{2}(\pi\ell(E)/d))\textnormal{d}E,
\end{equation}
where $\ell(E)$ is the oscillation length and $\theta_ {12}$ the solar mixing angle. This procedure was repeated for all reactors and the individual fluxes were summed for different detector locations. The necessary data concerning the coordinates was taken from the International Nuclear Safety Center (INSC) \cite{INSCwww} and includes all major power plants running worldwide in the year 2005. Future changes as the launch of the Finnish nuclear plant TVO3 as well as the possible shutdown of Swedish plants have not been taken into account but will not substantially change the $\bar\nu_e$ fluxes. The integral $\bar{\nu}_{e}$ fluxes and the corresponding event rates for LENA are summarized in Table\,\ref{nufluxes}. The calculated flux ($\phi=2.1\times10^6\,\text{cm}^{-2}\text{s}^{-1}$) in the Kamioka mine (Japan) is comparable to the one actually measured in KamLAND and is given for reference \cite{kamland-2003-1}. In addition, Fig.\,\ref{3reactors} shows the energy spectra of the reactor $\bar\nu_e$ at Frejus, Pyh\"asalmi and Hawaii, i.e. at a high, medium and low-flux site. \\
For an estimate of the overall uncertainties, we have taken into account possible deviations due to the fraction of annual runtime $e$ as well as the temporal variation of the abundancies of Uranium and Thorium isotopes ($\sim5\,\%$) \cite{fukugita}. In addition, the most recent values and uncertainties of the oscillation parameters $\Delta$m$_{12}^2=8.2^{+0.6}_{-0.5}\times 10^{-5}\,\text{eV}^2$ and $\tan^2\theta_{12}=0.40^{+0.10}_{-0.07}$ \cite{kamland-2004-1} have been included in the calculations. In Fig.\,\ref{3reactors}, these overall uncertainties are indicated as shaded regions.

\begin{table}
	\caption{Reactor $\bar\nu_e$ fluxes and event rates for LENA at various detector locations. Uncertainties of both fluxes and event rates are $\sim14\,\%$.}\label{nufluxes} 
	\begin{ruledtabular}
		\begin{tabular}{lccc}
		\multicolumn{2}{l}{}& reactor $\bar\nu_e$ flux  & reactor $\bar\nu_e$ events\\
		\multicolumn{2}{l}{Detector Location} & \scriptsize{(cm$^{-2}$s$^{-1}$)} & \scriptsize{(0.5 Mt yrs)}\\
		\hline
		Kamioka     & \scriptsize{(J)}  & $2.1 \times 10^6$ & $2.5 \times 10^5$ \\
  	Frejus      & \scriptsize{(F)}  & $1.6 \times 10^6$ & $2.0 \times 10^5$ \\
		Kimballton  & \scriptsize{(US)} & $6.4 \times 10^5$ & $7.3 \times 10^4$ \\
		Pyh\"asalmi & \scriptsize{(FIN)} & $1.9 \times 10^5$ & $2.1 \times 10^4$ \\
		Pylos       & \scriptsize{(GR)} & $9.2 \times 10^4$ & $11.5 \times 10^3$ \\
		Homestake   & \scriptsize{(US)} & $7.5 \times 10^4$ & $8.6 \times 10^3$ \\
		Henderson   & \scriptsize{(US)} & $7.4 \times 10^4$ & $8.4 \times 10^3$ \\
		Hawaii      & \scriptsize{(US)} & $10.9 \times 10^3$ & $12.4 \times 10^2$ \\
		Wellington  & \scriptsize{(NZ)} & $5.4 \times 10^3$ & $6.2 \times 10^2$ \\
		\end{tabular}
	\end{ruledtabular}
\end{table}

\subsection{\texorpdfstring{Atmospheric $\bar\nu_e$}{atmospheric barnue}}
\label{atm}

The flux of the atmospheric $\bar{\nu}_{e}$ is increasing with their energy and starts to surpass the DSN signal at energies around 25\,MeV. However, the total flux of atmospheric $\bar{\nu}_{e}$ is dependent on the geographic (geomagnetic) latitude and will therefore depend on the detector site \cite{agaisser-1988}. The individual spectra can be calculated using MC methods. However, for easier comparison and in accordance with publications by Ando and Sato we have extrapolated the model spectra calculated by Gaisser \textit{et al.} \cite{agaisser-1988} (including the corrections by Barr \textit{et al.} \cite{barr-1989}) to energies below 60\,MeV. We have estimated the atmospheric $\bar\nu_e$ spectra for high geomagnetic latitudes as, for instance, for Pyh\"asalmi using the 3D simulations by Liu \textit{et al.} \cite{Liu0211}.\\
As the energy spectrum of the atmospheric $\bar\nu_e$ only mildly depends on location \cite{agaisser-1988}, we left the spectrum unchanged and have only taken into account the dependence of the total flux on the detector site. Table\,\ref{atmtab} shows our results. The total flux of atmospheric $\bar\nu_e$ at a particular site can be related to the flux at the Kamioka site by a scaling factor $s_{atm}$.\\
In water Cherenkov detectors, atmospheric $\nu_\mu$ and $\bar\nu_\mu$ provide an additional background source by creating "invisible muons" \cite{ando-2003-1} with energies below the Cherenkov threshold. However, such muons do not pose a background in a liquid-scintillator detector due to its different detection mechanism.


\begin{table}
	\caption{Dependence of the total atmospheric neutrino flux below 60\,MeV on the detector location. The scaling factor $s_{atm}$ compares this flux to the one at the Kamioka site.}\label{atmtab}
	\begin{ruledtabular}
		\begin{tabular}{lll}
		Site & latitude (N) & $s_{atm}$  \\
		\hline
		Hawaii &	1.5$^{\circ}$ & 0.8 \\
		Kamioka, Pylos, Kimballton & 36.5$^{\circ}$, 36.6$^{\circ}$, 37.4$^{\circ}$ & 1 \\
		Henderson, Wellington & 39.8$^{\circ}$, -41.5$^{\circ}$ & 1.25 \\
		Frejus, Homestake & 45.1$^{\circ}$, 44.3$^{\circ}$  & 1.5 \\
		Pyh\"asalmi & 63.7$^{\circ}$ & 2.0 \\
		\end{tabular}
	\end{ruledtabular}	
\end{table}

\section{Cosmogenic Background}
\label{muons}

Up to now, the discussion only included background events due to additional $\bar\nu_e$ sources. However, muons that pass the detector or the surrounding medium (rock or water) also have to be considered. Most of the spallation products of these muons can be easily discriminated due to the signature of the inverse beta decay. Still, radionuclides like the $\beta$-$n$-emitter $^9$Li \cite{zbiri-06} or fast neutrons can mimic this $e^+-n$ coincidence. For an estimate of the rates, only events in the energy window from 10 to 25\,MeV have to be considered. All our calculations were performed for LENA at Pyh\"asalmi assuming a depth of 3960 m.w.e. and a corresponding muon flux of $(1.1\pm0.1)\times10^{-4}$/m$^2$s \cite{Enqvist0506}.

\subsection{\texorpdfstring{$^9$Li In-situ Production}{Li9 in-situ production}}
As the $\beta$ endpoint of $^9$Li is at 13.6\,MeV, it will affect the DSN detection only in the lower energy region of the observational window. For a rough estimate of the expected event rate, one can adopt the value derived for KamLAND: 0.6 events in 0.28~kt$\times$yrs exposure for $\text{E}>9.5$\,MeV \cite{kamland-2004}. Scaling the mass to 44~kt$\times$yrs fiducial volume in LENA, including a reduction of the integral muon flux by a factor of $\sim9$ in Pyh\"asalmi \cite{kudryavtsev-2003} and considering the dependence of the production rate on the muon energy ($\propto \text{E}_\mu^{0.75}$) \cite{thagner-1999}, the resulting rate is approximately 20 events per year.\\
However, as the muon passes through the fiducial volume, it can be clearly identified and a time as well as a volume cut can be applied. The $^9$Li cannot travel far, as its half-life is T$_{1/2}=0.18$\,s. Therefore, excluding a cylindrical volume of 2\,m around each muon's path for 1 second ($\sim5\times\text{T}_{1/2}$) decreases the background sufficiently. As the fiducial volume is hit by a muon about every 5\,s, one loses about $\sim 0.2\,\%$ of exposure time.

\subsection{Fast Neutrons}
Whereas muon tracks in the inner detector or in the muon veto can be clearly identified, muons passing the surrounding rock generate a background of fast neutrons. Usually, such a neutron will be accompanied by a number of charged particles that can be identified in the veto. However, there is a certain probability that the neutron will pass into the inner detector unnoticed. The neutron deposits its remaining kinetic energy in the scintillator and thereafter is captured by a proton, mimicking a real $\bar\nu_e$ event.\\
Using the estimates by Kudryavtsev et al. \cite{kudryavtsev-2003} for neutron production at 4000\,m.w.e. and assuming a mean absorption length of 0.75\,m \cite{kudryavtsev-2003}, one obtains a rate of about $10^5$ neutrons per year entering the muon veto. However, if one assumes a more realistic energy spectrum at the boundary between rock and cavern \cite{kudryavtsev-2003}, simulations performed with Geant4 \cite{geant4} show that only $\sim7$ neutrons per year will reach the fiducial volume and only 0.5 per year will generate a signal in the relevant energy region from 10 to 25\,MeV. As mentioned before, this number will be further decreased by the detection of accompanying shower particles in the muon veto and is therefore a conservative value. In the further discussion we assume the neutron spectrum to be energy independent.

\section{Detection Potential of LENA}
\label{discpot}

This Section mainly deals with the dependence of the observational window and the signal-to-background ratio on the actual detector location. As Pyh\"asalmi is one of the preferred sites, a statistical analysis has been carried out for this place in order to investigate the sensitivity of the detector regarding the separation of different DSN model predictions and the possibility to give constraints on the SNR. The results of such a spectral analysis are described in part \ref{analysis}.

\subsection{Energy Window}
\label{window}

Using the information on the two $\bar\nu_e$ background sources in a LSD (see Sect.\,\ref{reactor} and \ref{atm}), it is possible to define an optimal energy window for DSN observation.\\
As both the atmospheric and the reactor neutrino flux are dependent on the selected detector site, an individual upper and lower energy limit for DSN observation has to be set for every location. In a spectral analysis it is desirable to include as much of the DSN signal as possible, without increasing the background signal too much. We have chosen the lower (upper) energy thresholds at those energy values where the flux of the KRJ model ($f_{SN}=1$) begins (ends) to dominate in comparison to the total background flux (see Fig.\,\ref{Pyhasalmi} described below). The site-dependent detection thresholds are listed in Table\,\ref{ewindow}. The best limits can be achieved at Hawaii with a window from about 8.4 to 29.0\,MeV. At Pyh\"asalmi, it will be from around 9.7 to 25.1\,MeV.

\begin{table}
	\caption{Lower and upper energy thresholds for DSN observation and predicted signal and background events inside the energy window for LENA at different locations after 10 years of measuring time and $f_{SN}=1$ (see Eq.\,(\ref{SFRf})). The range in the signal rates is determined by the LL and TBP model predictions (\textit{no\,resonance}, see Sect.\,\ref{theory}). The background includes atmospheric and reactor $\bar\nu_e$ events and signals due to fast neutrons.}\label{ewindow}
	\begin{ruledtabular}
		\begin{tabular}{lcrr}
		\multicolumn{2}{l}{} & Energy Window & DSN S/B\\
		\multicolumn{2}{l}{Detector Location} & \scriptsize{(MeV)} & \scriptsize{(0.5\,Mt\,yrs)} \\
		\hline
		Kamioka     & \scriptsize{(J)}  &  11.1 - 28.1 & 21-42/11\\
  	Frejus      & \scriptsize{(F)}  &  10.8 - 26.4 & 22-41/12\\
		Kimballton  & \scriptsize{(US)} &  10.6 - 28.1 & 23-44/11\\
		Pyh\"asalmi & \scriptsize{(FIN)}&  9.7 - 25.1  & 24-45/13\\
		Pylos       & \scriptsize{(GR)} &  9.4 - 28.1  & 27-49/12\\
		Homestake   & \scriptsize{(US)} &  9.0 - 26.4  & 28-49/13\\
		Henderson   & \scriptsize{(US)} &  8.9 - 27.2  & 28-50/13\\
		Hawaii      & \scriptsize{(US)} &  8.4 - 29.0  & 31-54/12\\
		Wellington  & \scriptsize{(NZ)} &  8.2 - 27.2  & 31-53/12\\
		\end{tabular}
	\end{ruledtabular}
\end{table}

\subsection{Event Rates}

\begin{figure}
  \begin{center}
    \includegraphics[width=0.45\textwidth]{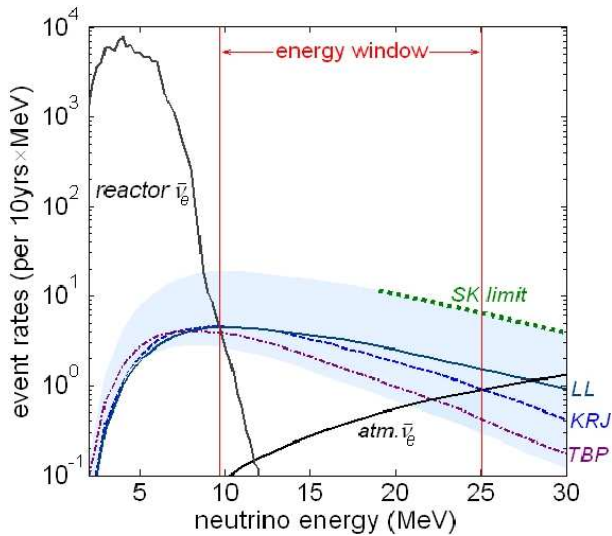}
    \caption{Event rates of reactor, atmospheric and DSN $\bar{\nu}_{e}$ (LL, KRJ, TBP, see Fig.\,\ref{SRNspec}) as expected for LENA in Pyh\"asalmi after ten years of measurement and $f_{SN}=1$. The shaded region represents the uncertainties of the DSN rates due to $f_{SN}$. The energy window is chosen such that the flux of the KRJ model exceeds the background flux. The Super-Kamiokande limit is also indicated.\label{Pyhasalmi}} 
  \end{center}	
\end{figure}

The detection thresholds as well as the event rates vary with the location that is chosen for LENA. As TBP provides the lowest event rates and LL the highest, their rates are given in Table\,\ref{ewindow} for the different detector sites. In addition, the number of background events inside the energy window is shown.\\
At Pyh\"asalmi, we have obtained between 24\,$f_{SN}$ (TBP) to 45\,$f_{SN}$ (LL) events in 10 years of measurement within the energy window (\textit{no\, resonance}, see Sect.\,\ref{theory}). In case of LL\textit{res}, 53\,$f_{SN}$ events would be detected. For a value of $f_{SN}=2.5$ \cite{strigari05}, one expects therefore $\sim100$ events in this time period. The background due to reactor and atmospheric neutrinos within the same time would give $\sim8$ events, and up to 5 events from fast neutrons have to be added. Taking into account the overall uncertainty in the supernova rate for $z=0$ (factor $f_{SN}=0.7-4.1$, see Eq. (\ref{SFRf})) the predicted DSN event range is widened further to $17-220$ events. However, the direct upper limit on the DSN flux of Super-Kamiokande \cite{SK-2003-SRNlimit} corresponds to 185 events in LENA. In any case, a significant detection signal of the DSN appears to be certain. The spectra of both signal and background events are shown in Fig.\,\ref{Pyhasalmi} for LENA located at Pyh\"asalmi.\\

It should be emphasized that in Pyh\"asalmi about 25\,$\%$ of the registered DSN in the energy window are originating from redshifts $z>1$. Their flux is expected to influence the spectral form at low energies visibly.\\

\textbf{Flux limits.} Assuming there is no signal of the DSN, we calculated the flux limit that LENA would be able to give for two energy regions: One for the most sensitive region from 10.5 to 19.3\,MeV where the background levels are lowest. A second one in the regime from 19.3\,MeV to 25\,MeV that corresponds to the limit achieved by the Super-Kamiokande detector \cite{SK-2003-SRNlimit}. Assuming a 100\,$\%$ efficiency for the detection of all $\bar\nu_e$ and other background events, one can derive the statistical uncertainty corresponding to 90\,$\%$ C.L.. If one assumes the spectral form of the KRJ model, one can match the number of DSN events to the uncertainty of the background by scaling the DSN flux by means of the factor $f_{SN}$. The derived limits on the flux are well below the current model predictions. Table\,\ref{limits} shows background event rates, flux limits after 10 years of measuring and lowest model predictions  for both energy regions. The limits of 0.3\,cm$^{-2}$s$^{-1}$ (0.13\,cm$^{-2}$s$^{-1}$) obtained by LENA are 20$\,\%$ (80$\,\%$) of the flux for $f_{SN}=0.7$ below (above) 19\,MeV. In addition, the limit from Super-Kamiokande is shown. LENA could improve the present limit by a factor of 9 within 10 years.

\begin{table}
	\caption{Using the calculated background levels in two energy bins, an upper limit on the DSN flux can be achieved. Values are given for 10 years of measuring and 100$\%$ detection efficiency both for signal and background. For comparison, lowest model predictions ($f_{SN}=0.7$ in the SNR) and the current Super-Kamiokande limit are also given.}\label{limits}
	\begin{ruledtabular}
		\begin{tabular}{lccl}
		Energy Range (MeV)			      & 10.5-19.3	& 19.3-25	& 									\\
		\hline
		atmosperic $\bar\nu_e$ events &	2.2		& 3.6 		&	in 10\,yrs				\\
		reactor $\bar\nu_e$ events 		& 0.2		& 0				&						  			\\
		fast neutron events						& 2.7		& 2.0			&						  			\\
		\hline
		total								          & 5.1	  & 5.6	    & in 10\,yrs				\\
		\hline
		flux limit (90\,$\%$ C.L.) 		& 0.3  	& 0.13	  & cm$^{-2}$s$^{-1}$	\\
		lowest model prediction				& 1.4  	& 0.16    & 	          			\\
		Super-Kamiokande limit				&       & 1.2			&										\\
		\end{tabular}
	\end{ruledtabular}	
\end{table}

\subsection{Spectral Analysis}
\label{analysis}

According to the current DSN model predictions, LENA will provide a sufficient event rate for probing both parameters that enter into the actual DSN spectrum: the redshift-dependent supernova rate (SNR) and the core-collapse SN $\bar\nu_e$ spectrum.

\begin{figure}
  \begin{center}
    \includegraphics[width=0.42\textwidth]{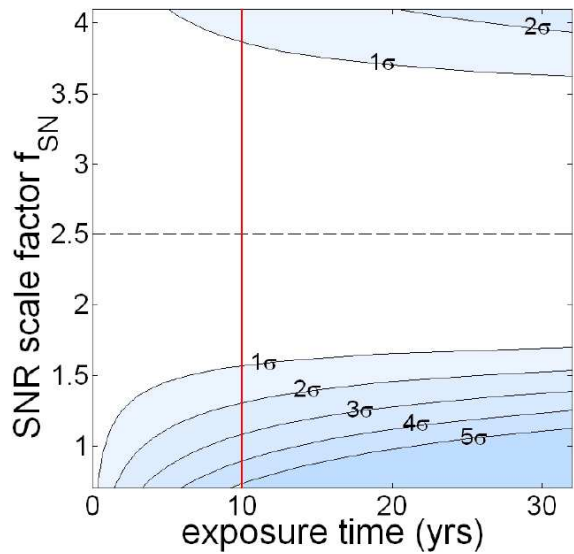}
    \caption{Potential to derive bounds on the scale factor $f_{SN}$ in the SNR. If one assumes $f_{SN}=2.5$ according to the CMSFR, the curves shown in the plot depict the significance level at which other values of $f_{SN}$ can be rejected after a certain exposure time.\label{SFRsep}} 
  \end{center}	
\end{figure}

\textbf{Limits on SNR.} The event numbers in the energy regime from 10 to 14\,MeV are quite similar for all DSN models, yielding between 1.48\,$f_{SN}$ to 1.97\,$f_{SN}$ events per year, as can be seen in Table\,\ref{MC}. Using the actually detected event rate in this energy bin, to derive limits on the scaling-factor $f_{SN}$ in the SNR the SN neutrino spectrum does not have to be known. Fig.\,\ref{SFRsep} shows LENA's exclusion potential if one assumes $f_{SN}=2.5$ to be true. The depicted curves show the significance level for rejection of other $f_{SN}$ values dependent on the exposure. After 10 years of measurement, $f_{SN}\leq1.3$ could be excluded at a 2$\sigma$ level.\\
In this way, LENA will be able to independently verify the value of $f_{SN}$ derived from optical observations of the SNR, and cross-check the validity of the applied dust corrections. At present, SNR observations reach out to $z\approx0.9$ \cite{lunardini0509,cappellaro9904,dahlen0406}. With the lower energy threshold of 9.7\,MeV for LENA in Pyh\"asalmi, there will be also a non-neglegible amount of DSN events due to SN at $1<z<2$ that can be used to derive limits on the SNR for this redshift region. Combined with assumptions about the IMF \cite{hopkins0601}, DSN observation will also be able to provide constraints on the SFR.


\textbf{Discrimination of DSN models.} The progress in optical observations is likely to provide a solid prediction for both the SFR and SNR up to $z\approx2$. Using this information, an analysis of the DSN event spectrum in LENA could be used to constrain the parameter range for different SN models or even to discriminate between them.\\
For testing LENA's potential, MC simulations of typical DSN event spectra have been carried out. In these tests, one DSN model and the currently favoured value of $f_{SN}=2.5$ have been assumed to be true, and in accordance with these assumptions 10$^4$ MC spectra have been simulated. As the event numbers are low, only two energy bins have been chosen: the first one reaching from 9.7 to 14.5\,MeV and the second one from 14.5 to 25.1\,MeV. In Table\,\ref{MC}, the expected event numbers per year for the three DSN models and the background rates are shown, including the \textit{resonance} case for the LL model (see Sect.\,\ref{theory}). As only two energy bins have been chosen, the excellent energy resolution of the detector of better than 3\,$\%$ for $\text{E}_{\bar\nu_e}>10$\,MeV and the reaction kinematics have almost no effect on the analysis. Both have therefore been neglected.

\begin{table}
	\caption{Rates for the DSN models used in the MC-calculations and $\chi^2$-tests performed in order to analyse the potential for model 
	discrimination. Values are given for 50\,kt\,yrs of exposure (one year of measuring time in LENA) at Pyh\"asalmi and $f_{SN}=2.5$.}\label{MC}
	\begin{ruledtabular}
	\begin{tabular}{lccc}
						  	& Rates ($N_{B1}$) & Rates ($N_{B2}$) \\
		DSN model 	& \scriptsize{9.7-14.5\,MeV} & \scriptsize{14.5-25.1\,MeV} \\
		\hline
		LL~~~~~\textit{(no\,res.)} &	4.93 & 6.20 \\
		TBP					& 3.70 & 2.65 \\
		KRJ					& 4.88 & 4.73  \\
		\hline
		LL~~~~~\textit{(res.)} & 4.50 & 8.70\\
		\hline
		background	& 0.60 & 0.73 \\
		\end{tabular}
	\end{ruledtabular}
\end{table}

\begin{figure}
  \begin{center}
    \includegraphics[width=0.48\textwidth]{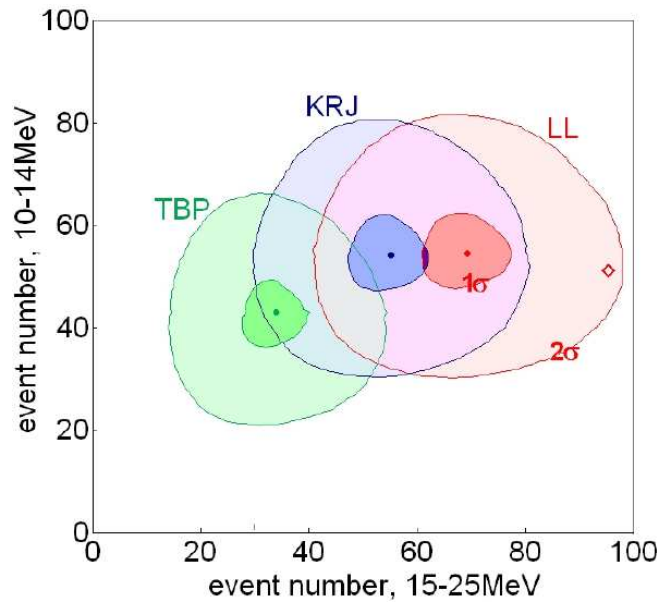}
    \caption{Exclusion plot for the assignment of a simulated event spectrum in LENA to a wrong DSN model. A value of $f_{SN}=2.5$ and 10 years of exposure are assumed. MC spectra created according to one of the models are compared to all possible combinations of event numbers in the two energy bins (see Tab.\,\ref{MC}). Regions of more than $1\sigma$ and $2\sigma$ exclusion probability for a wrong assignment are located outside the depicted lines. Predictions assuming the \textit{no\,resonance} (\textit{dots}) and the \textit{resonance} case (\textit{diamond}) are shown. LL and TBP model can be discerned at a significance level of more than $2.6\sigma$. \label{sep}} 
  \end{center}	
\end{figure}

Via a $\chi^2$-analysis, an exclusion probability can be given that a MC spectrum created according to a given model is wrongly assigned to a different combination of event numbers and therefore to an alternative SN model. Fig.\,\ref{sep} shows exclusion plots for each of the three models assuming an exposure time of 10 years. $2\sigma$ and $1\sigma$ exclusion regions are located outside the depicted curves. For comparison, the predicted event numbers for all models are shown.\\
The evaluation shows that the separation potential for the models LL and TBP is best, as their spectral slopes differ most. For $f_{SN}$=2.5, LENA is able to  discriminate between these two models with a significance of 2.6$\sigma$ after 10 years of exposure. However, a separation at 2$\sigma$ level between spectra with very similar slopes as KRJ and LL would require long exposure times of $>30$ years unless $f_{SN}$ is very close to the upper limit ($f_{SN}\leq4.1$). For the case of the LL model, the exclusion probability for the other models is shown in Table\,\ref{excl} as a function of the exposure time. Note that an exclusion at $>2\sigma$ of a resonant flavour conversion in the SN envelope (corresponding to LL\textit{res}) would be possible after 15 years of measuring.

\begin{table}
	\caption{Exclusion probability for a wrong model assignment of a simulated DSN event spectrum in LENA. The LL model (\textit{no\,resonance} case) is assumed to be true, significance levels of the rejection of other models are shown as a function of the exposure time and $f_{SN}$.}\label{excl}
	\begin{ruledtabular}
	\begin{tabular}{ccccc}
		\multicolumn{2}{l}{Parameters} & \multicolumn{3}{c}{Significance of exclusion ($\sigma$)} \\
		\hline
		$f_{SN}$  & exposure (yrs) & TBP & KRJ & LL\textit{res} \\
		\hline
		2.5				& 5 			& 1.8	& 1.1 &	1.5 \\
							& 10			& 2.6	& 1.3 &	1.8 \\
							& 15			& 3.0 &	1.5	& 2.1 \\
							& 20			& 3.3	& 1.6	& 2.4 \\
							& 25			& 3.7	& 1.8	& 2.6 \\
							& 30			& $>$4 & 1.9 & 2.9 \\
	  \hline
	  4.1				& 30			& $>$4 & 2.3 & 3.6 \\
	\end{tabular}
	\end{ruledtabular}
\end{table}

\section{Conclusions}
\label{conclusions}

It has been shown in this paper, that in a large liquid-scintillator detector (LSD) like LENA diffuse supernova neutrino (DSN) events can be detected in an almost background-free energy window between $\sim10$ and 25\,MeV. High background suppression is achieved due to the fact that the neutron produced in the detection reaction - the inverse beta decay - can be detected in a LSD. Above $\sim10$\,MeV the high-energetic tail of the man-made reactor $\bar\nu_e$ is becoming negligible in comparison to the DSN flux. Above $\sim25$\,MeV the atmospheric $\bar\nu_e$ flux starts to dominate.\\
If placed in Pyh\"asalmi, the lower threshold for LENA will be close to 9.7\,MeV. For this location and for the most likely value of $f_{SN}=2.5$ (see Eq.\,(\ref{SFRf})) in the supernova rate (SNR), about 6 to 13 events per year will be contained in the energy window and can be detected. This would be the first detection of the DSN background. It should also be noted that about 25\,$\%$ of the detected DSN will be originating from a red-shift region $1<z<2$. If no signal was detected, a new limit of 0.13\,cm$^{-2}$s$^{-1}$ on the flux above 19.3\,MeV could be achieved within 10 years that would surpass the one of the Super-Kamiokande detector by a factor $\sim9$. In the lowest background region between 10.5 and 19.3\,MeV the limit of 0.3\,cm$^{-2}$s$^{-1}$ would be 20$\,\%$ of the lowest current model predictions.\\
Apart from mere detection, LENA will be able to distinguish between different DSN models and give constraints on the form of the neutrino spectrum emitted by a core-collapse supernova. This can be reached via an analysis of the DSN's spectral slope. The significance of the results will be highly dependent on the exposure time and the supernova rate in the near universe. For a known SNR with $f_{SN}=2.5$, the discrimination between the discussed LL and TBP models for the DSN will be possible at a 2.6$\sigma$ level after 10 years of measuring time. Distinguishing between DSN models with more similar spectral slopes, however, would require higher statistics.\\
In addition, by an analysis of the flux in the energy region from 10 to 14\,MeV the SFR for $z<2$ could be constrained at high significance levels. The current lower bound of $f_{SN}=0.7$ \cite{ando0401} could be increased to $f_{SN}=1.3$ at a 2$\sigma$ level within 10 years of measuring time.\\
LENA will therefore not only be able to detect the DSN; it will also be able to make valuable contributions to both, core-collapse SN models and the redshift dependent supernova rate.

\section*{Acknowledgements}
We thank G. Raffelt, T. Janka and S. Ando for both valuable information and helpful suggestions concerning supernova physics and diffuse supernova neutrinos. This work has been supported by funds of the Maier-Leibnitz-Laboratorium (Garching), of the Deutsche Forschungsgemeinschaft DFG (Sonderforschungsbereich 375), and of the Virtual Institute for Dark Matter And Neutrinos (VIDMAN).\\
\vspace{13cm}


\bibliographystyle{h-physrev}
\bibliography{dsn}

\begin{thebibliography}{10}

\bibitem{ando0401}
{S. Ando and K. Sato},
\newblock New J. Phys. {\bf 6}, 170 (2004), astro-ph/0410061 v2.

\bibitem{sk-dsn-06}
C.~Lunardini,
\newblock Phys. Rev. D {\bf 73}, 083009 (2006), hep-ex/0601054.

\bibitem{sfr-05}
L.~Strigari {\em et~al.},
\newblock JCAP {\bf 0504}, 017 (2005), astro-ph/0502150.

\bibitem{lunardini0610}
C.~Lunardini, 2006, astro-ph/0610534.

\bibitem{SK-2003-SRNlimit}
Super-Kamiokande collaboration, M.~S. Malek {\em et~al.},
\newblock Phys. Rev. Lett. {\bf 90}, 061101 (2003), hep-ex/0209028.

\bibitem{kamland-2004}
K.~Eguchi {\em et~al.},
\newblock Phys. Rev. Lett. {\bf 92}, 071301 (2004), hep-ex/0310047.

\bibitem{borex-sn}
L.~Cadonati,
\newblock Astropart.Phys. {\bf 16}, 361 (2002).

\bibitem{sno+dsnb}
M.~Chen,
\newblock Talk at the Conference on Neutrino Geophysics, Honolulu, Hawaii,
  December 15, 2005.

\bibitem{ando-2003-18}
S.~Ando {\em et~al.},
\newblock Astroparticle Physics {\bf 18}, 307 (2003), astro-ph/0202450.

\bibitem{LENA-05}
L.~Oberauer {\em et~al.},
\newblock Nucl. Phys. B (Proc. Suppl.) {\bf 138}, 108 (2005).

\bibitem{LENA-06}
{T. Marrod\'an Undagoitia} {\em et~al.},
\newblock Prog. Part. Nucl. Phys. {\bf 57}, 283 (2006), hep-ph/0605229.

\bibitem{borexcoll-2004-1}
BOREXINO collaboration, S.~Sch\"onert {\em et~al.}, 2004, physics/0408032,
\newblock submitted to NIM A.

\bibitem{mydipl}
M.~Wurm,
\newblock Diploma Thesis, Technische Universit\"at M\"unchen, Germany, 2005.

\bibitem{hochmuth05}
K.~A. Hochmuth {\em et~al.}, 2005, hep-ph/0509136.

\bibitem{Oberauer0402}
L.~Oberauer,
\newblock Mod.Phys.Lett. A {\bf 19}, 337 (2004), hep-ph/0402162.

\bibitem{Skadhauge0611}
S.~Skadhauge and R.~Z. Funchal,
\newblock (2006), hep-ph/0611194.

\bibitem{teresa-2005-1}
T.~{Marrod\'an Undagoitia} {\em et~al.},
\newblock Phys. Rev. D {\bf 72}, 075014 (2005), hep-ph/0511230.

\bibitem{Marrodan06TAUP}
T.~{Marrod\'an Undagoitia} {\em et~al.},
\newblock J. Phys.: Conf. Ser. {\bf 39}, 269 (2006).

\bibitem{peltoniemi0506}
J.~Peltoniemi,
\newblock Talk at the 7th International Workshop on Neutrino Factories and
  Superbeams, Frascati (Rome), Italy, June 21, 2005.

\bibitem{ando0402}
S.~Ando,
\newblock Astrophys. J. {\bf 607}, 20 (2004), astro-ph/0401531.

\bibitem{LL-1998}
T.~Totani {\em et~al.},
\newblock Astrophys. J. {\bf 496}, 216 (1998).

\bibitem{TBP-2003}
T.~A. Thompson, A.~Burrows, and P.~Pinto,
\newblock Astrophys. J. {\bf 592}, 434 (2003), astro-ph/0211194.

\bibitem{KRJ-2003}
M.~T. Keil, G.~G. Raffelt, and H.~T. Janka,
\newblock Astrophys. J. {\bf 590}, 971 (2003), astro-ph/0208035.

\bibitem{dighe9907}
A.~Dighe and A.~Smirnov,
\newblock Phys. Rev. D {\bf 62}, 033007 (2000), hep-ph/9907423.

\bibitem{Kotake0509}
K.~Kotake {\em et~al.},
\newblock Rep. Prog. Phys. {\bf 69}, 971 (2006), astro-ph/0509456.

\bibitem{Chooz9711}
{CHOOZ Collaboration},
\newblock Phys. Lett. B {\bf 420}, 397 (1998), hep-ex/9711002.

\bibitem{cappellaro9904}
E.~Cappellaro {\em et~al.},
\newblock Astron.Astrophys. {\bf 351}, 459 (1999), astro-ph/9904225.

\bibitem{dahlen0406}
T.~Dahlen {\em et~al.},
\newblock Astrophys.J. {\bf 613}, 189 (2004), astro-ph/0406547.

\bibitem{lunardini0509}
C.~Lunardini,
\newblock (2005), astro-ph/0509233.

\bibitem{hopkins0601}
A.~Hopkins and J.~F. Beacom,
\newblock (2006), astro-ph/0601463.

\bibitem{UV-1}
S.~Lilly {\em et~al.},
\newblock Astrophys. J. {\bf 460}, L1 (1996).

\bibitem{UV-2}
P.~Madau {\em et~al.},
\newblock Mon. Not. R. Astron. Soc. {\bf 283}, 1288 (1996).

\bibitem{UV-3}
C.~C. Steidel {\em et~al.},
\newblock Astrophys. J. {\bf 519}, 1 (1999).

\bibitem{FIR-1}
D.~H. Hughes {\em et~al.},
\newblock Nature {\bf 394}, 241 (1998).

\bibitem{FIR-2}
H.~Flores {\em et~al.},
\newblock Astrophys. J. {\bf 517}, 148 (1999).

\bibitem{strigari05}
L.~E. Strigari {\em et~al.},
\newblock JCAP {\bf 0504}, 017 (2005), astro-ph/0411424.

\bibitem{Strumia0302}
A.~Strumia and F.~Vissani,
\newblock Phys.Lett. B {\bf 564}, 42 (2003), astro-ph/0302055.

\bibitem{tengblad-1989}
O.~Tengblad {\em et~al.},
\newblock Nucl. Phys. A {\bf 503}, 136 (1989).

\bibitem{fukugita}
{M. Fukugita and T. Yanagida},
\newblock {\em Physics of Neutrinos and Applications to Astrophysics} (Springer
  Verlag, Berlin, 2003).

\bibitem{LBNLwww}
{LBNL Isotopes Project},
\newblock http://ie.lbl.gov/toi.htm.

\bibitem{letourneau-2005-talk}
{A. Letourneau and D. Lhuillier},
\newblock private communication.

\bibitem{Zacek86}
G.~Zacek {\em et~al.},
\newblock Phys. Rev. D {\bf 34}, 2621 (1986).

\bibitem{kopeikin-2003}
V.~I. Kopeikin {\em et~al.},
\newblock Phys. Atom. Nucl. {\bf 67}, 1892 (2004), hep-ph/0410100.

\bibitem{IAEAwww}
{International Atomic Energy Agency},
\newblock http://www.iaea.org.

\bibitem{INSCwww}
{International Nuclear Safety Center},
\newblock http://www.insc.anl.gov.

\bibitem{kamland-2003-1}
KamLAND collaboration, K.~Eguchi {\em et~al.},
\newblock Phys. Rev. Lett. {\bf 90}, 021802 (2003).

\bibitem{kamland-2004-1}
KamLAND collaboration, T.~Araki {\em et~al.},
\newblock Phys. Rev. Lett. {\bf 94}, 081801 (2004), hep-ex/0406035 v3.

\bibitem{agaisser-1988}
T.~K. Gaisser {\em et~al.},
\newblock Phys. Rev. D {\bf 38}, 85 (1988).

\bibitem{barr-1989}
G.~Barr {\em et~al.},
\newblock Phys. Rev. D {\bf 39}, 3532 (1989).

\bibitem{Liu0211}
Y.~Liu {\em et~al.},
\newblock Phys. Rev. D {\bf 67}, 073022 (2003), astro-ph/0211632.

\bibitem{ando-2003-1}
S.~Ando,
\newblock Phys. Lett. B {\bf 570}, 11 (2003), hep-ph/0307169.

\bibitem{zbiri-06}
K.~Zbiri {\em et~al.},
\newblock (2006), hep-ph/0607179 v2.

\bibitem{Enqvist0506}
T.~Enqvist {\em et~al.},
\newblock NIM A {\bf 554}, 286 (2005), hep-ex/0506032.

\bibitem{kudryavtsev-2003}
V.~A. Kudryavtsev {\em et~al.},
\newblock Nucl. Instrum. Meth. A {\bf 505}, 688 (2003), hep-ex/0303007.

\bibitem{thagner-1999}
T.~Hagner {\em et~al.},
\newblock Astropart. Phys. {\bf 14}, 33 (2000).

\bibitem{geant4}
{S. Agostinelli $et~al.$ (Geant4)},
\newblock Nucl. Instrum. Meth. Phys. Res., Sect. A {\bf 506}, 250 (2003).

\end{thebibliography}

\end{document}